# Hydrophobic-hydrophilic interactions drive rapid nanoparticle assembly

Benjamin F. Porter[1], Mercè Pacios[1] and Harish Bhaskaran[1*]

[1] Department of Materials, University of Oxford, Parks Road, Oxford OX1 3PH (United Kingdom).

*E-mail: harish.bhaskaran@materials.ox.ac.uk

**The unique physical and optical properties of nanoparticles make their reliable and rapid assembly an important goal for nanotechnology. To this end, many competing driving forces have been targeted to control the self-assembly, though issues with reliability and efficiency indicate that interfaces on the nanoscale are not yet fully understood. In this letter we report for the first time that the flow of water media from the interface between chemically nanopatterned hydrophobic surfaces and adjacent hydrophilic regions (a Janus interface) can effectively be employed for rapid nanoparticle assembly with single nanoparticle resolution. We find that the dewetting of water molecules from this Janus interface drives the assembly of nanoparticles from electrostatically stabilised nanoparticle colloids onto the hydrophobic areas, expedited by removing the resistance to adhesion between wetted interfaces exerted by any hydration barrier, such that nanoparticle assembly is achieved in under a minute. We call this method of controlled nanoparticle assembly "Janus Interface Nanoparticle Assembly - JINA". We then demonstrate control of this mechanism through pH switching and prove its effectiveness for fabrication by rapidly creating single-nanoparticle sites and single particle wide lines whilst simultaneously densely coating macro-scale features. Not only does this assembly method have far-reaching implications in nanomanufacturing, the unanticipated nature of this mechanism also challenges many commonly held assumptions about the interfacial interactions of nanoparticle self-assembly.**

Efforts to use the bottom-up fabrication of nanoscale components into complex devices hinges on the ability to engineer systems that create a desired final state by self-assembly[1,2]. Assembling nanoscale components into pre-defined geometries has been demonstrated using electrostatic[3–6], dielectrophoretic[7,8], magnetic, DNA[9–11] and geometry[12–19] driven processes with promising applications in a number of fields. Capillary assembly can achieve impressive, dense assemblies of nanoscale objects into geometrically defined templates[16,19] and DNA self-assembly can be made directional in 3D by customising the orientation of bonding regions on colloidal surfaces[10]. Self-assembly of nanoscale components by electrostatic interactions is intuitive, attracting and repelling oppositely and like charged components respectively. These processes typically require timescales on the order of hours to achieve the desired assembly. The long-term effectiveness of functional and charged monolayers is also lowered by the decomposition of the active ligands or interference from other interactions.

Hydrophobic surfaces are known to exhibit strong attractive forces between each other in polar solvents such as water[20,21]. A hydrophobic force law accounting for all experimental conditions is an ongoing objective of theoretical research[22], but is generally considered to involve the dewetting of water molecules between the interfaces, as water molecules are entropically more stable in the bulk or on hydrophilic surfaces than on a hydrophobic one. Careful control of the composition of non-polar and polar solvents has even enabled hydrophobic nanomaterials to be reversibly assembled into larger clusters[23,24]. In the case of competing hydrophobic and hydrophilic interactions at a Janus interface, it has generally been considered that hydrophilic interactions are stronger and prevent dewetting[21,25]. Simulations of the dewetting at a Janus interface have indicated that the

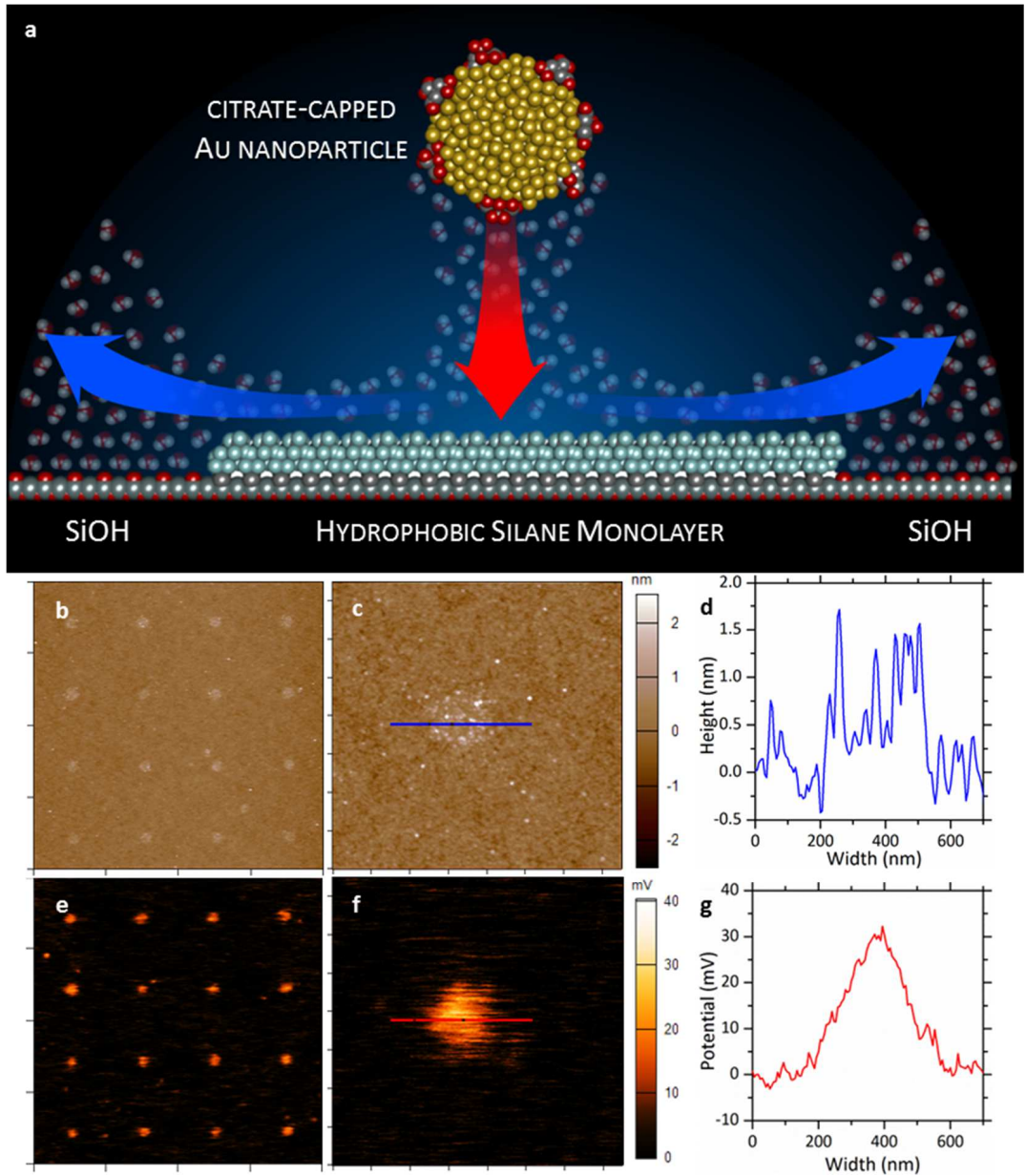

**Figure 1 Concept and characterisation of substrates for assembly using JINA**

(a) Conceptual image of the JINA process: a 20 nm citrate-stabilised Au nanoparticle is driven towards a hydrophobic surface by the rapid dewetting of the water from the hydrophobic to the surrounding hydrophilic, hydroxylated silica surface. The hydrophobicity of the surface destabilises the hydration barrier on the nanoparticle causing rapid adhesion. Mutual electrostatic repulsion drives particles away from the silica surface and areas already occupied by nanoparticles. (b, c) Atomic Force and (e, f) Kelvin Probe Microscopy maps of 250 nm diameter POTS monolayers patterned on the $SiO_2$. (b, e) are 8 x 8 μm scans of 16 POTS areas in a 4 x 4 array with 2 μm pitch and (c, f) are 1.2 x 1.2 μm scans of a single 250 nm diameter POTS areas. (d, f) are lines scans corresponding to the blue and red lines in (c) and (f) respectively.

dewetting of the surface could occur, but only if the hydrophilic surface carried a sufficiently small surface charge[26].

In this paper, we introduce a new assembly technique that we call JINA – Janus Interface Nanoparticle Assembly – driven by the dewetting of water from nanoscale patterned hydrophobic monolayers to adjacent hydrophilic regions, enabling the rapid self-assembly of nanoparticles onto the hydrophobic areas. A conceptual image of this assembly is presented in Figure 1a, and is animated in Supplementary Video 1. The surfaces for the assembly are hydrocarbon or fluorocarbon molecular monolayers functionalized onto a masked silica surface. In contact with the water-based Au colloidal suspension, the hydroxylated silica surface adopts a negative surface charge ($SiOH + OH^- \leftrightarrow SiO^- + H_2O$) that is repulsive to the negatively charged citrate-capping layer on the Au nanoparticles. Water molecules form hydration barriers at these charged surfaces, with the polarity of the molecules aligned to the surface charge. The alignment of these barriers is also repulsive between like-charged surfaces.

In contrast, at the hydrophobic surface the water molecules are entropically more stable in the bulk or the surrounding hydroxylated silica, so the water dewets from this interface. As the molecules diffuse away from the hydrophobic surface to the surrounding charged silica and back into the bulk, a pressure gradient driving nanoparticles towards the surface is created as illustrated in Figure 1a. When the nanoparticle and hydrophobic surfaces are in close proximity, the remaining water rapidly dewets from this Janus interface and the particle is driven into contact with the surface. Although previous experimental measurements of such a Janus interface showed that the hydration barrier would remain stable[25], the dynamics of the unrestrained nanoparticle colloid in our experiments have proven the hydrophobic interaction can also dominate.

To test this mechanism, we fabricated devices by electron-beam lithography using 100 nm thick 495K molecular weight poly (methyl methacrylate) (PMMA) layer, spin-coated on silicon substrates with a 100 nm thick, thermally-grown $SiO_2$ surface. We functionalized hydrophobic molecular monolayers onto the exposed oxide areas by immersion in a 5 mM solution of trimethoxy(octadecyl)silane (TMODS) in a propan-2-ol solvent. After complete growth of the monolayer, thorough rinsing and removal of the PMMA resist left devices patterned with hydrophobic monolayers and unfunctionalised oxide. Using amplitude-modulated Atomic Force Microscopy (AFM) we measured the topography of these monolayers on the substrates (Figure 1b, c), finding them to have a thickness of 1.3 ± 0.6 nm (Figure 1d). Crucially, the contrast in surface potential measured by Kelvin Probe Microscopy (KPM) in parallel to the topography measurements (Figure 1e-g) confirms that this is a different surface material to the oxide surface, verifying that the functionalization had taken place (KPM resolution limited by the 100 nm thick oxide). Further verification that the lithography pattern was preserved during self-assembly of the monolayer is presented in Supplementary Information S1.

Self-assembly using the proposed JINA mechanism was then attempted with colloidal suspensions of 20 nm Au particles (with $7.00 \times 10^{11}$ particles per mL) that are stabilised by negatively charged citrate capping ions. We carried this out by depositing a colloidal droplet (100 μL) on top of the patterned area and allowing for assembly to occur spontaneously. After rinsing in deionised water and drying in a stream of nitrogen gas, we observed that the nanoparticles had self-assembled into an ordered layer on the hydrophobic surfaces. Our initial tests with the hydrocarbon TMODS monolayer achieved assembly of the nanoparticles after 1 minute (Figure 2a) and dense assembly of the nanoparticles after 4 minutes (Figure 2b). A supplementary AFM topography image of these densely assembled nanoparticles on

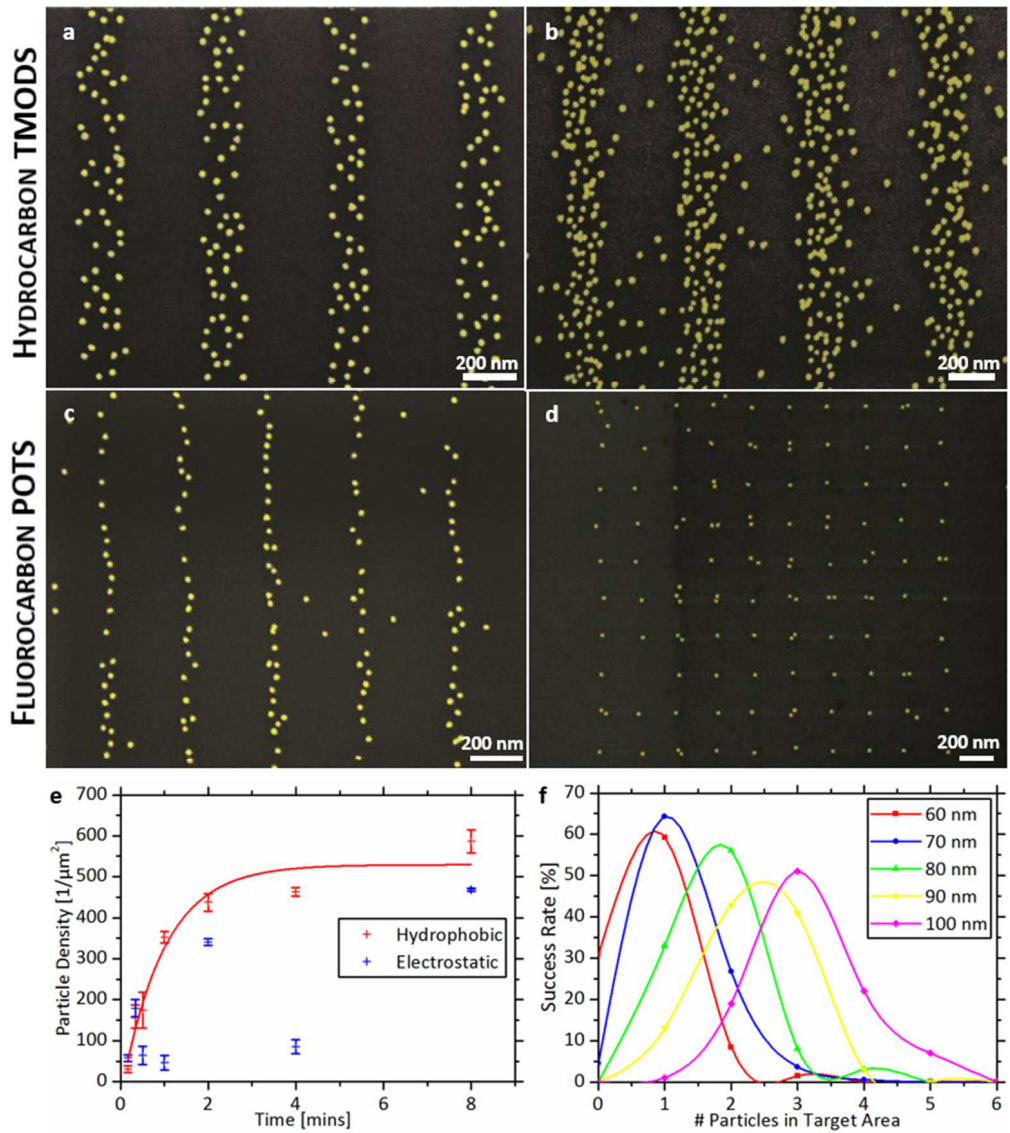

**Figure 2 Nanoparticles self-assembled onto hydrophobic monolayers**

(a, b) Assembly of the 20 nm Au nanoparticles onto 200 nm wide bands of hydrophobic TMODS after (a) 1 minute and (b) 4 minutes of exposure to the Au colloid. (c, d) Assembly of the nanoparticles onto SiO₂ surfaces patterned with a hydrophobic POTS monolayer. (c) Assembly approaching single particle width on POTS lines 50 nm in width. (d) 10 x 10 array of 70 nm circular POTS surfaces with 250 nm pitch. Individual particles were captured on the majority of these target areas. (e) A chart of the increase in particle density accrued on 500 nm wide POTS bands with increasing deposition time. Comparison to equivalent electrostatically attractive surfaces shows frequent instability of electrostatic assembly on short timescales. (f) A chart of the number of nanoparticles deposited onto POTS areas with diameters ranging from 60 to 100 nm after 1 minute of assembly, showing the highest statistical chance of single particle placement for diameters between 60 and 70 nm.

TMODS is presented in Figure S2. This showed particles had quickly assembled onto the surface, taking only a few minutes to achieve dense surface coatings. We also observe that the electrostatic repulsions between the nanoparticles in the colloid are not disrupted by JINA, as the particles have assembled separately onto the surface.

To achieve a stronger hydrophobic interaction, the hydrocarbon-based TMODS was replaced with 1H,1H,2H,2H-perfluorooctyltriethoxysilane (POTS) to take advantage of the greater hydrophobicity of fluorocarbons[27]. Results of these experiments using the same 20 nm Au particles are presented in the Scanning Electron Microscopy (SEM) images in Figure 2c and 2d, following 1 minute of particle assembly. Using POTS in place of TMODS resulted in a greater resolution of particle self-assembly, with lines of particles approaching one nanoparticle width deposited on 60 nm wide lines of the hydrophobic monolayer (Figure 2c). Further analysis of the particle density accrued on hydrophobic POTS surfaces with time (Figure 2e) indicated that the particle density on areas 500 nm in width would increase rapidly during the first 2 minutes before plateauing and reaching a high packing density. For comparison, the same study was performed on patterned, electrostatically attractive monolayers of (3-aminopropyl) triethoxysilane (APTES). With the electrostatic monolayer, results appeared to be less stable than through JINA, with particles frequently detaching from the surface to spread to the surrounding oxide surface. Those that did not experience this exhibited a similar assembly on the surface, though of a markedly lower particle density. Further results of using JINA with different particle sizes and concentrations are provided in Supplementary Information S3.

Achieving reliable control of single nanoparticles on a large scale is a nontrivial problem. Using these fluorocarbon monolayers for JINA achieved consistent single nanoparticle placement after just 1 minute on arrays of 60 to 70 nm diameter circular POTS areas (Figure 2d). To determine its effectiveness, we analysed the results for different diameters of POTS monolayers, looking at 400 target areas for each diameter of POTS considered and compared the results (Figure 2f). Comparing the results to our goal of achieving single particle placement (SPP), the best results were achieved for a diameter of 70 nm, with 64.3 % of target regions achieving SPP after just 1 minute of assembly.

For applications in additive nanomanufacturing, it is very beneficial if a fabrication process can work over both macro and nanoscales simultaneously. To demonstrate the capability of the JINA process to this aim, we attempted to create macro scale assembly of an arbitrary pattern with single particle detail as well as multiparticle areas (we chose the iconic London skyline). The result of this experiment is presented in the colourized SEM images in Figure 3 where, as before, 20 nm Au particles were assembled by JINA in 1 minute onto patterned monolayers. In these figures, it is clear that the large features have been densely coated with nanoparticles (Figure 3a), whilst the details of the London Eye (Figure 3b-d, which are zooms into the high resolution regions within the Ferris Wheel of Figure 3a) have achieved nanoscale detail with particles assembled into single particle wide lines.

The driving mechanism of the JINA process we have described so far relies upon the dewetting of the hydration barrier on the nanoparticle surface when it approaches the hydrophobic layer. If this were indeed the case as we have hypothesised, when the surface charge and thus the barrier are made stronger, then the dewetting and thereby the particle deposition should be prevented[26]. This can be achieved by adjusting the pH of the colloid, which would change the surface charge of the Au nanoparticles; the citrate ions that

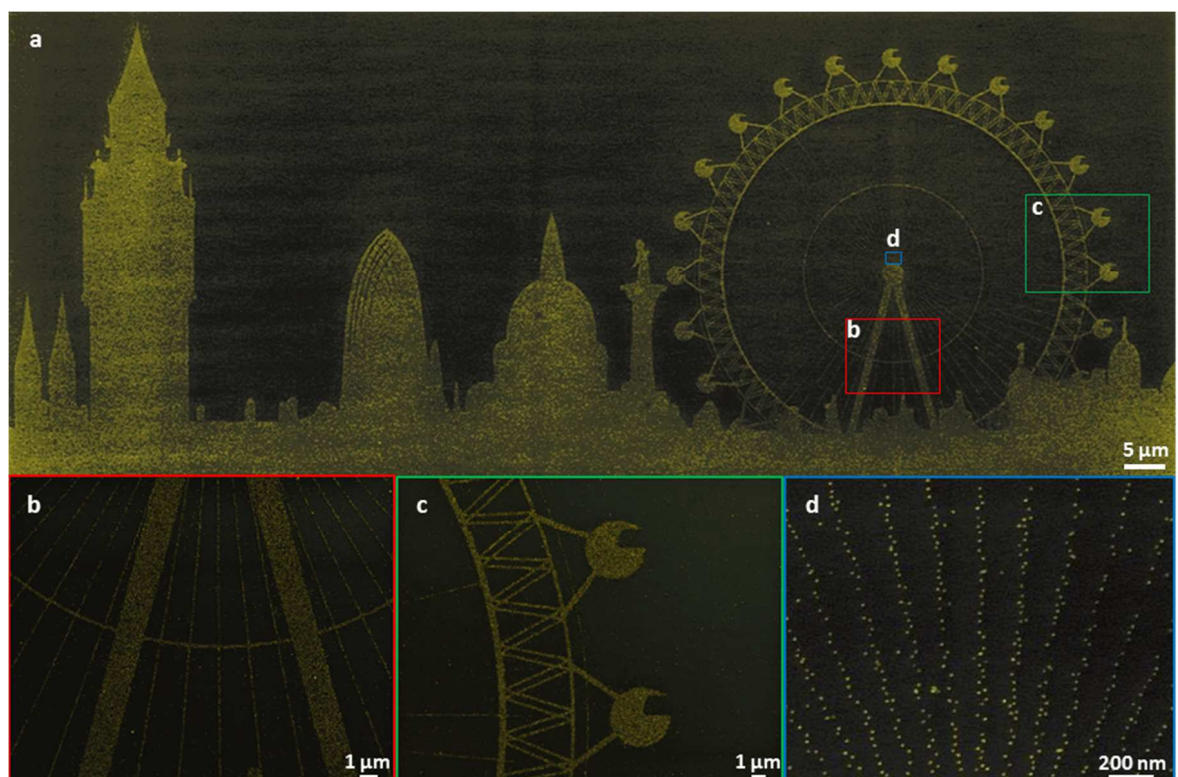

**Figure 3 Single-step large area and single nanoparticle-wide assembly of the London Skyline by JINA**

Colourized SEM images showing the simultaneous multi- and single nanoparticle assembly of particles by JINA. (a) The full 200 μm London skyline formed by the assembly of 20 nm Au nanoparticles onto patterned POTS monolayers after 60 s. Dense coating of nanoparticles were assembled on areas > 10 μm in size by the same process as the single nanoparticle-wide spokes of "The London Eye" (b-d) Zoom into relevant regions of matching, coloured border in (a) demonstrating single particle resolution, which is achieved simultaneously and many times very close to multi-particle assembly.

form the stabilising, capping layer for these particles are multivalent, with pKa values of 3.13, 4.76 and 6.40. The initial pH of the 20 nm colloid was measured at 6.8, so the citrate ions should be fully ionised. However, the diffuse double layer that forms around charged surfaces in solution will contain a greater proportion of cations with increasing pH, resulting in further increases in the zeta potential with pH[28].

To test whether the assembly could be suppressed this way, we prepared colloids with a range of pH values via addition of HCl and NaOH and observed their assembly after 5

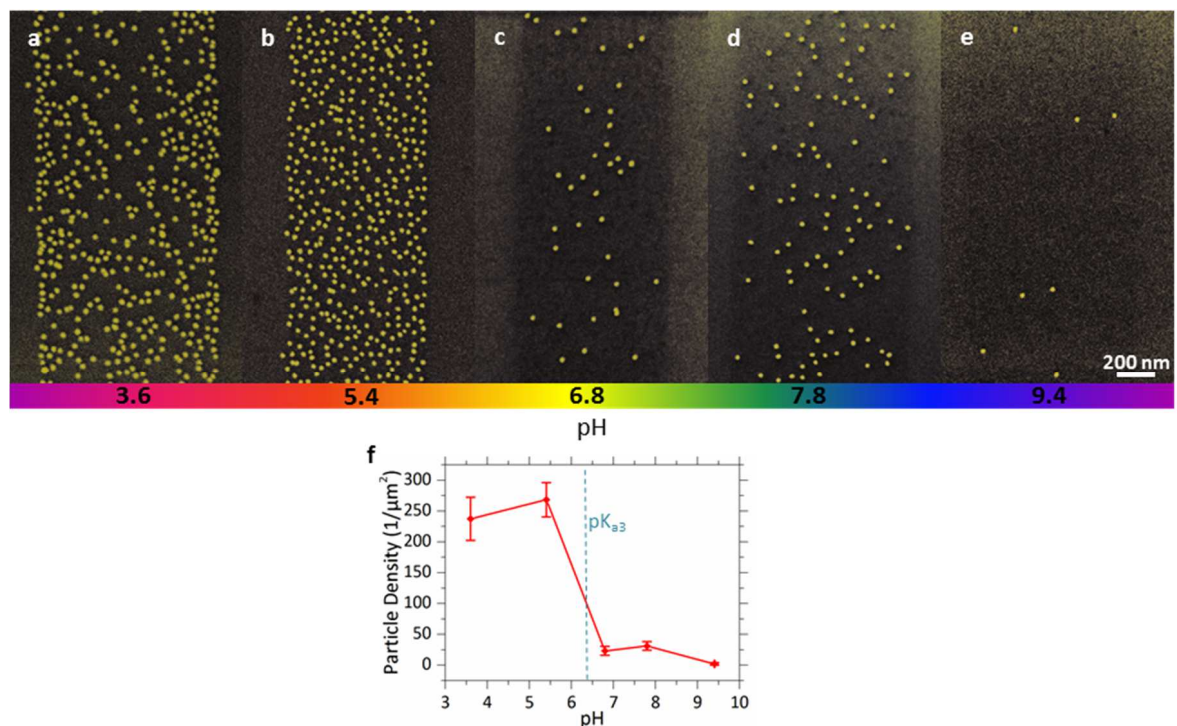

**Figure 4 pH dependence of hydrophobic-hydrophilic self-assembly**

Colourized SEM images showing the effect that pH adjustment had on the hydrophobic self-assembly process. After 5 minutes the 20 nm Au colloid is seen to self-assemble onto a 700 nm wide band of TMODS down the centre of each image, with the pH adjusted to (a) 3.6, (b) 5.4, (c) 7.8 and (d) 9.4. The monolayer is distinguished from the background when fewer nanoparticles assembled onto the surface at higher pH. The scale bar in these SEM images is 200 nm. (f) The trend between the density of particles assembled on the the hydrophobic monolayer and the pH of the colloid used. The third acid dissociation constant $pK_{a3}$ of the citrate capping ions is marked, showing that the assembly was significantly increased by a reduction in the colloid surface charge across this threshold.

minutes onto patterned lines of TMODS. The colloid was diluted to a particle concentration of $1.4 \times 10^{11}$ per mL (pH remained at 6.8) to reduce the risk of destabilising the colloid. An ionic strength above 1 mM was found to destabilise this colloid, so the pH range of colloids investigated was limited to between 3 and 11. At a pH of 3.6 (Figure 4a) we observed disordered assembly of particles onto the TMODS. The colloid itself had become unstable as was evidenced by the faded colour of the dispersion indicating significant agglomeration of particles. This is compared to the slightly denser and more ordered assembly observed at a pH of 5.4 where the colloid was more stable (Figure 4b). At higher pH levels (Figure 4d, e) the assembly of particles on the hydrophobic surface is greatly reduced and can almost be suppressed completely. The overall trend of decreasing assembled particle density with increasing pH is clearly seen in Figure 4f (error bars are standard deviation between different measurements across the surface). This shows that the surface charge of the Au nanoparticles was sufficient to prevent dewetting when the pH was increased, inhibiting the deposition of nanoparticles onto the TMODS surface. Conversely, the deposition was denser when the pH and thus particle charge were decreased, which is consistent with simulations of Janus interfaces[26]. This pH response indicates that there is a balance to be established between maintaining a stable colloid and reducing the pH to promote particle assembly. Crucially, it also points to a possible method to not only control assembly by changing pH, but potentially allowing controlled disassembly in liquid.

Another important aspect of this self-assembly is that it occurs rapidly. The widespread use of self-assembly methods in manufacturing is hindered by the speed at which it can achieve effective results. What we found in our experiments was that the assembly of particles through JINA into single nanoparticle arrays required only a few seconds to achieve this. In Figure 5 we see how the results after 20 s, 30 s, 120 s and 240 s successfully capture very similar numbers of nanoparticles on the hydrophobic surfaces in each case. Importantly, this contrasts with the time dependent deposition we had observed for larger areas > 100 nm in Figure 2f. This implies that the dewetting between the hydrophobic/hydrophilic interface has a stronger effect when the size of the hydrophobic region is <100 nm. This is consistent with our explanation that the assembly occurs by the flow induced at the Janus Interface, as

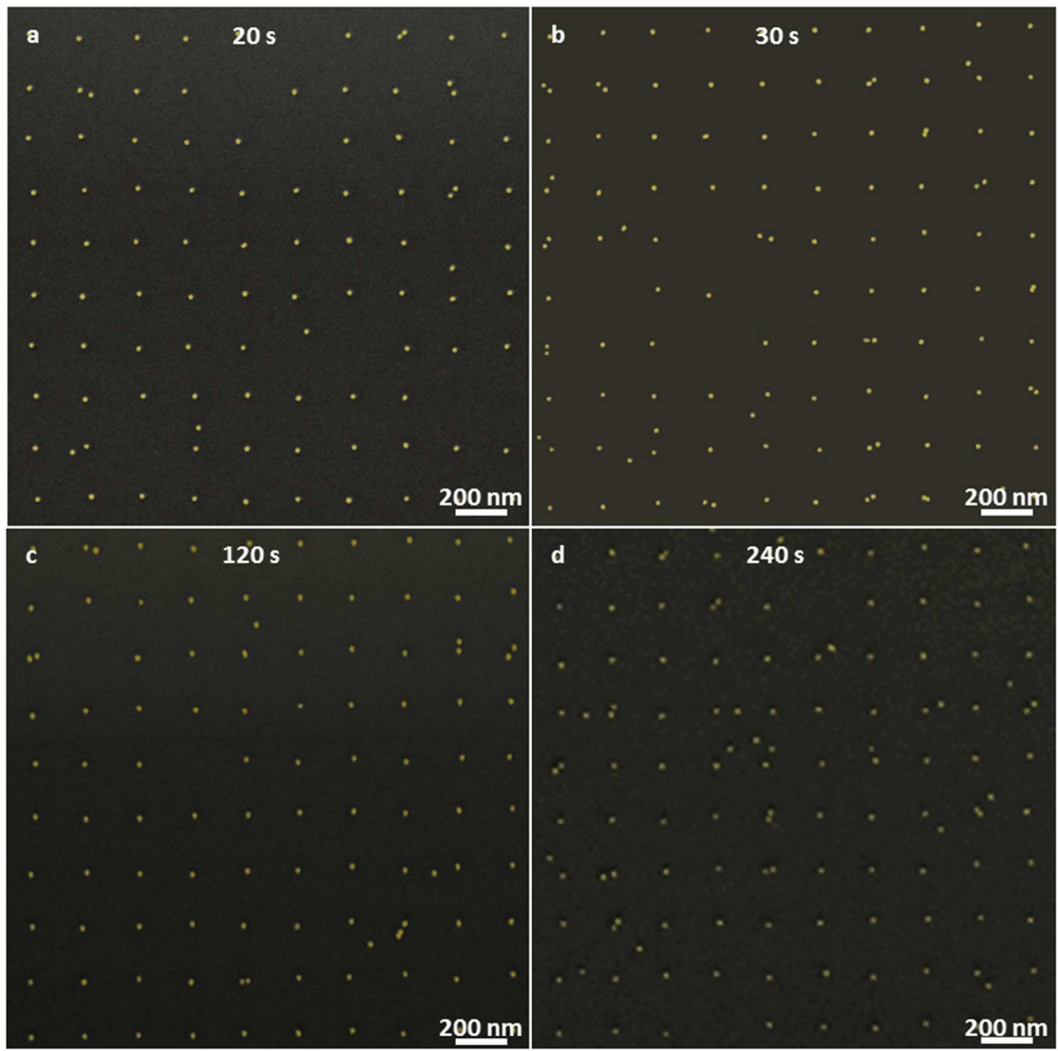

**Figure 5 Speed of single-particle resolution self-assembly**

SEM images showing the short time needed for the assembly of arrays of single Au nanoparticles to occur on hydrophobic POTS monolayers. The number of particles captured in these arrays is very similar for deposition in (a) 20 s, (b) 30 s, (c) 120 s and (d) 240 s. This indicates that assembly onto the hydrophobic surface will occur within seconds with little improvement on nanoscale features over longer timescales.

flow would be more intense in regions of higher pressure gradient, an effect that is size dependent.

The stray particles visible in the assembly on our devices appears to be more determinant on the destabilisation of particles during rinsing than any dependence on the time given for deposition to occur for nanoscale features like these. The dewetting of the hydration barrier at this Janus interface and subsequent assembly is thus a very rapid process with a high level of precision that could lead to the integration of self-assembly into widespread use in nanomanufacturing. Further improvements could be made by enhancing the hydrophilicity of the surrounding $SiO_2$ surface to prevent destabilisation of the barrier.

One important feature of the JINA process is that it will not occur if there is no hydrophobic-hydrophilic interaction present on the surface. For example, with a surface that was completely hydrophobic (monolayer grown without patterning) we observed that the hydrophobic surface showed no significant particle assembly over similar time scales to previous results. The JINA mechanism thus hinges on the ability of water molecules to move to hydrophilic regions adjacent to hydrophobic patterns that draw the colloid droplet into contact with the patterned features, thus validating the Janus Interface hypothesis. Additionally, the particle assembly does not occur if a hydrophobic monolayer is not grown in the fabrication process – any charging of the surface from the electron beam lithography therefore does not drive the particle assembly, ruling out accidental electrostatic assembly due to electron beam charging. We also tested whether nanoparticles would become readily detached by immersing and agitating them in high pH solutions as well as non-polar solvents did not induce a detachment of the particles from the surface into the surrounding medium. Such a mechanism may only be possible without drying the substrates after the particles have assembled.

A simple electrokinetic quantification of the TMODS and POTS surfaces indicate that the hydrophobic surfaces would confer a strong, negative zeta potential (approximately -30 to -50 mV) at the pH range of the colloid used[29,30]. This would suggest that the negatively charged gold nanoparticles would not undergo the particle assembly we have observed. Surface energies are low for hydrophobic coatings presenting a low barrier to particle adhesion, but this alone does not explain why particles did not assemble onto surfaces coated completely with a hydrophobic monolayer, nor the similar though less reliable assembly of particles onto electrostatically attractive amine surfaces (Figure 2e). Clearly, the complex interactions between all the media on this nanoscale, the highly transient nature of hydrodynamic phenomena that can occur on the nanoscale[31] and numerous other parameters (local pH and ionic concentration/compositions, temperature, precise surface chemistry, etc.) are all involved in a fairly complex interaction that is beyond the scope of this article; further investigation will be required to understand these complex interactions. However, the experiments we have described in this paper supports our hypothesis to explain the rapid assembly of nanoparticles we have observed.

In conclusion, we have engineered a Janus Interface Nanoparticle Assembly (JINA) mechanism, whereby Au nanoparticles undergo rapid self-assembly onto lithographically defined hydrophobic/hydrophilic interfaces, achieving single nanoparticle resolution in a matter of seconds. We have surmised that this is due to rapid dewetting of the hydration barrier to the proximate hydrophilic surroundings, which we demonstrated can be controlled through manipulation of the surface charge via pH switching. The control afforded by this process and the single particle resolution of JINA are potentially crucial for the development of nanoassembly replicators to rapidly construct complex 3D nanotechnologies that exploit the unique properties of nanoparticles[32]. The use of nanoparticles in nanomedicine could also be influenced by our findings, due to the prevalence of hydrophobic and hydrophilic media in biological environments. The nature of the JINA mechanism observed indicates that fast nanoparticle assembly does not solely depend on the interaction of the ligands of the active materials; rather how these ligands interact with the intervening medium plays a dominant role. The full impact that these findings could have for various fields of research will become known as future efforts investigate the occurrence of this mechanism in other colloids and Janus interfaces.

## Methods

**Device fabrication:** 10 x 10 mm Si chips with a thermally grown 100 nm polished oxide surface (purchased from IDB Technologies Ltd.) were coated with a 100 nm of PMMA by spin coating at 4000 rpm for 60 s (3 s ramp up and down) and baking at $180^0$C for 90 s. Electron beam lithography with a 50 kV Jeol 5500F tool patterned the surface with a charge density of 325 $\mu$C/cm$^2$ followed by development for 33 s in a 15:5:1 solution of propan-2-ol, methyl isobutyl ketone and ethyl methyl ketone and blanched in propan-2-ol for 15 s followed by rinsing, drying and a post-bake of $120^0$C for 120 s. Patterned samples were immersed in a 5 mM solution of trimethoxyocatadecylsilane or 1H,1H,2H,2H-perfluorooctyltriethoxysilane (as applicable) in IPA for 24 hours for monolayer growth. Samples were dried with compressed $N_2$ and rinsed thoroughly with propan-2-ol before undergoing lift-off by sonication in acetone. Dried samples were left in a vacuum desiccator for at least 24 hours prior to testing.

**Device Testing:** Au colloids were sonicated for 5 minutes prior to deposition to reduce aggregation. 100 $\mu$L of the colloid was deposited on top of the patterned surface of the sample and left to assemble for a given amount of time. The sample was then rinsed in two rounds of deionised water for 30 s each and a final rinse in propan-2-ol for 60 s and drying with compressed $N_2$.

**Sample characterisation**: Atomic Force and Kelvin Probe Microscopy (AFM/KPM) were performed with an Asylum MFP-3D AFM, using 81.2 kHz silicon PPP-FMR Nanosensors AFM probes. Scanning Electron Microscopy (SEM) was performed at 10 kV with a Jeol 6500F and a Hitachi S-4300 SEM. Analysis of particle density after assembly was performed using ImageJ[33].

## Acknowledgements

This research was supported via an EPSRC Manufacturing Fellowship EP/J018694/1, the WAFT Collaboration (EP/M015173/1) and the OUP John Fell Fund. We are grateful to A. Ne for useful scientific discussions.

## Contributions

BFP and MP carried out the experimental work and analyses. HB led the research and with BFP, wrote the manuscript. All authors contributed significantly to this work.

## Additional information

Supplementary information is available. Correspondence and requests for materials and devices should be addressed to H.B.

## Competing financial interests

The authors declare no competing financial interests.

# References

1. Engstrom, D. S., Porter, B., Pacios, M. & Bhaskaran, H. Additive nanomanufacturing – A review. *J. Mater. Res.* **29,** 1792–1816 (2014).

2. Min, Y., Akbulut, M., Kristiansen, K., Golan, Y. & Israelachvili, J. The role of interparticle and external forces in nanoparticle assembly. *Nat. Mater.* **7,** 527–538 (2008).

3. Ma, L.-C. *et al.* Electrostatic funneling for precise nanoparticle placement: a route to wafer-scale integration. *Nano Lett.* **7,** 439–45 (2007).

4. Huang, H.-W., Bhadrachalam, P., Ray, V. & Koh, S. J. Single-particle placement via self-limiting electrostatic gating. *Appl. Phys. Lett.* **93,** 73110 (2008).

5.  Ray, V. *et al.* CMOS-compatible fabrication of room-temperature single-electron devices. *Nat. Nanotechnol.* **3,** 603–608 (2008).

6.  Jiang, M., Kurvits, J. a, Lu, Y., Nurmikko, A. V. & Zia, R. Reusable Inorganic Templates for Electrostatic Self-Assembly of Individual Quantum Dots, Nanodiamonds, and Lanthanide-Doped Nanoparticles. *Nano Lett.* **15,** 5010–5016 (2015).

7.  Padmaraj, D. *et al.* Parallel and orthogonal E-field alignment of single-walled carbon nanotubes by ac dielectrophoresis. *Nanotechnology* **20,** 35201 (2009).

8.  Xiong, X. *et al.* Directed assembly of gold nanoparticle nanowires and networks for nanodevices. *Appl. Phys. Lett.* **91,** 63101 (2007).

9.  Zheng, Y. *et al.* Gutenberg-style printing of self-assembled nanoparticle arrays: electrostatic nanoparticle immobilization and DNA-mediated transfer. *Angew. Chem. Int. Ed. Engl.* **50,** 4398–4402 (2011).

10. Wang, Y. *et al.* Colloids with valence and specific directional bonding. *Nature* **490,** 51–55 (2012).

11. Rogers, W. B. & Manoharan, V. N. Programming colloidal phase transitions with DNA strand displacement. *Science (80-. ).* **347,** 639–642 (2015).

12. Asbahi, M. *et al.* Large Area Directed Self-Assembly of Sub-10 nm Particles with Single Particle Positioning Resolution. *Nano Lett.* 150814114502001 (2015). doi:10.1021/acs.nanolett.5b02291

13. Asbahi, M. *et al.* Template-Induced Structure Transition in Sub-10 nm Self-Assembling Nanoparticles. *Nano Lett.* **14,** 2642–2646 (2014).

14. Morton, S. W. *et al.* Scalable manufacture of built-to-order nanomedicine: spray-assisted layer-by-layer functionalization of PRINT nanoparticles. *Adv. Mater.* **25,** 4707–13 (2013).

15.    Duan, H. & Berggren, K. K. Directed self-assembly at the 10 nm scale by using capillary force-induced nanocohesion. *Nano Lett.* **10,** 3710–3716 (2010).

16.    Kraus, T. *et al.* Nanoparticle printing with single-particle resolution. *Nat. Nanotechnol.* **2,** 570–576 (2007).

17.    Holzner, F. *et al.* Directed Placement of Gold Nanorods Using a Removable Template for Guided Assembly. *Nano Lett.* **11,** 3957–3962 (2011).

18.    Krishnan, M., Mojarad, N., Kukura, P. & Sandoghdar, V. Geometry-induced electrostatic trapping of nanometric objects in a fluid. *Nature* **467,** 692–696 (2010).

19.    Flauraud, V. *et al.* Nanoscale topographical control of capillary assembly of nanoparticles. *Nat. Nanotechnol.* **12,** 73–81 (2016).

20.    Israelachvili, J. & Pashley, R. The hydrophobic interaction is long range, decaying exponentially with distance. *Nature* **300,** 341–342 (1982).

21.    Israelachvili, J. N. *Intermolecular and Surface Forces with Applications to Colloidal and Biological Systems*. (Academic Press Inc. (London) Ltd., 1985).

22.    Hammer, M. U., Anderson, T. H., Chaimovich, A., Shell, M. S. & Israelachvili, J. N. The search for the hydrophobic force law. *Faraday Discuss.* **146,** 299–308 (2010).

23.    Sánchez-Iglesias, A. *et al.* Hydrophobic interactions modulate self-assembly of nanoparticles. *ACS Nano* **6,** 11059–11065 (2012).

24.    Nie, Z. *et al.* Self-assembly of metal–polymer analogues of amphiphilic triblock copolymers. *Nat. Mater.* **6,** 609–614 (2007).

25.    Zhang, X., Zhu, Y. & Granick, S. Hydrophobicity at a Janus interface. *Science* **295,** 663–666 (2002).

26.    Hua, L., Zangi, R. & Berne, B. J. Hydrophobic Interactions and Dewetting between Plates with Hydrophobic and Hydrophilic Domains. *J. Phys. Chem. C* **113,** 5244–5253

(2009).

27.    Dalvi, V. H. & Rossky, P. J. Molecular origins of fluorocarbon hydrophobicity. *Proc. Natl. Acad. Sci.* **107,** 13603–13607 (2010).

28.    Shyue, J. J. *et al.* Acid-base properties and zeta potentials of self-assembled monolayers obtained via in situ transformations. *Langmuir* **20,** 8693–8698 (2004).

29.    Lin, C. & Chaudhury, M. K. Using Electrocapillarity to Measure the Zeta Potential of a Planar Hydrophobic Surface in Contact with Water and Nonionic Surfactant Solutions. *Langmuir* 14276–14281 (2008).

30.    Pazokifard, S., Mirabedini, S. M., Esfandeh, M. & Farrokhpay, S. Fluoroalkylsilane treatment of TiO2 nanoparticles in difference pH values: Characterization and mechanism. *Adv. Powder Technol.* **23,** 428–436 (2012).

31.    Tandon, V. & Kirby, B. J. Zeta potential and electroosmotic mobility in microfluidic devices fabricated from hydrophobic polymers: 2. Slip and interfacial water structure. *Electrophoresis* **29,** 1102–1114 (2008).

32.    Daniel, M. C. M. & Astruc, D. Gold Nanoparticles: Assembly, Supramolecular Chemistry, Quantum-Size Related Properties and Applications toward Biology, Catalysis and Nanotechnology,. *Chem. Rev.* **104,** 293–346 (2004).

33.    Schneider, C. A., Rasband, W. S. & Eliceiri, K. W. NIH Image to ImageJ: 25 years of image analysis. *Nat. Methods* **9,** 671–675 (2012).